\documentclass[journal=jacsat,manuscript=article]{achemso}
\usepackage{multirow}
\usepackage{amsmath}
\usepackage{graphicx}
\usepackage{bm}
\usepackage{braket}
\usepackage{soul} 
\usepackage[normalem]{ulem}

\setstcolor{red}
\usepackage[dvipsnames]{xcolor}

\newcommand{\MC}[1]{\textcolor{black}{#1}}

\newcommand{\wh}[1]{\textcolor{black}{#1}}
\newcommand{\QO}[1]{\textcolor{black}{#1}}

\author{Wenfei Li}
\altaffiliation{Contributed equally to this work}
\affiliation{AI for Science Institute, Beijing 100080, P.R.~China}
\author{Qi Ou}
\altaffiliation{Contributed equally to this work}
\affiliation{AI for Science Institute, Beijing 100080, P.R.~China}
\author{Yixiao Chen}
\affiliation{Program in Applied and Computational Mathematics, Princeton University, Princeton, NJ 08544, USA}
\author{Yu Cao}
\affiliation{HEDPS, CAPT, College of Engineering and School of Physics, Peking University, Beijing 100871, P.R.~China}
\author{Renxi Liu} 
\affiliation{HEDPS, CAPT, College of Engineering and School of Physics, Peking University, Beijing 100871, P.R.~China}
\author{Chunyi Zhang}
\affiliation{Department of Physics, Temple University, Philadelphia, PA 19122, USA}
\author{Daye Zheng}
\affiliation{AI for Science Institute, Beijing 100080, P.R.~China}
\author{Chun Cai}
\affiliation{AI for Science Institute, Beijing 100080, P.R.~China}
\altaffiliation{DP Technology, Beijing 100080, P.R.~China}
\author{Xifan Wu}
\affiliation{Department of Physics, Temple University, Philadelphia, PA 19122, USA}
\author{Han Wang}
\email{wang_han@iapcm.ac.cn}
\affiliation{Laboratory of Computational Physics, Institute of Applied Physics and Computational Mathematics, Huayuan Road 6, Beijing 100088, P.R.~China}
\author{Mohan Chen}
\email{mohanchen@pku.edu.cn}
\affiliation{HEDPS, CAPT, College of Engineering and School of Physics, Peking University, Beijing 100871, P.R.~China}
\author{Linfeng Zhang}
\affiliation{AI for Science Institute, Beijing 100080, P.R.~China}
\altaffiliation{DP Technology, Beijing 100080, P.R.~China}
\email{linfeng.zhang.zlf@gmail.com}

\title{DeePKS+ABACUS as a Bridge between Expensive Quantum Mechanical Models and Machine Learning Potentials} 

\begin{document}




\begin{abstract}
Recently, the development of machine learning (ML) potentials has made it possible to perform large-scale and long-time molecular simulations with the accuracy of quantum mechanical (QM) models.
However, for high-level QM methods, such as density functional theory (DFT) at the meta-GGA level and/or with exact exchange, quantum Monte Carlo, etc., generating a sufficient amount of data for training a ML potential has remained computationally challenging due to their high cost.  
In this work, we demonstrate that this issue can be largely alleviated with Deep Kohn-Sham (DeePKS), a ML-based DFT model.
DeePKS employs a computationally efficient neural network-based functional model to construct a correction term added upon a cheap DFT model.
Upon training, DeePKS offers closely-matched energies and forces compared with high-level QM method, but the number of training data required is orders of magnitude less than that required for training a reliable ML potential.
As such, DeePKS can serve as a bridge between expensive QM models and ML potentials: one can generate a decent amount of high-accuracy QM data to train a DeePKS model, and then use the DeePKS model to label a much larger amount of configurations to train a ML potential.
This scheme for periodic systems is implemented in a DFT package ABACUS, which is open-source and ready for use in various applications.
\end{abstract}

\section{Introduction}
Over the past few decades, rapid developments of high speed and massively parallel computing have boosted the exploration of tremendous microscopic phenomena in condensed phases. 
For such investigations, one of the most widely applied tools is molecular dynamics (MD), which models atomic and molecular systems by numerically solving the Newtonian equations of motion subject to specific boundary conditions.\cite{jcp_md1,jcp_md2} 
The interatomic energies and forces involved in the Newtonian equations can be either obtained via an empirical force field (EFF),\cite{eff1,eff2,eff3} or computed from $ab\ initio$ calculations, known as $ab\ initio$ MD (AIMD).\cite{car1985unified,marx_hutter_2009,carloni_aimd,cornell1995second,del2013bulk} 
Despite the high efficiency of EFF-based MD, its applications are sometimes inhibited due to less satisfying modeled results compared to experiments as well as the presumably questionable transferability when employed to a new system.\cite{mackerell2004empirical,vellore2010assessment} 

AIMD simulation generates trajectories by performing quantum mechanical (QM) calculations ``on-the-fly'' as the simulation proceeds. 
For condensed systems, density functional theory (DFT)\cite{kohn1965self} is usually the QM method of choice for AIMD owing to its relatively balanced treatment for the trade-off between efficiency and accuracy. 
In the framework of DFT, the performance of AIMD simulations rests on the selection of the DFT exchange-correlation (XC) functionals. 
Liquid water, for example, cannot be quantitatively modeled by AIMD with normal general gradient approximation (GGA) functionals\cite{perdew1996generalized}, which lacks proper description of van der Waals (vdW) interactions.\cite{distasio2014individual,mohan_pnas} 
Going beyond GGA, meta-GGA or even hybrid meta-GGA functionals that lie on higher rungs of Perdew’s metaphorical Jacob's ladder offer significantly more accurate predictions, albeit with manifold increased computational cost.\cite{mohan_pnas,chunyi_jpcb} 
Therefore, AIMD simulations with those higher-rung functionals are inevitably limited to fairly small systems with a short simulation time scale (tens of picoseconds), which prevents the quantitative investigation on macroscopic properties of which the converged predictions require much longer simulation time and larger system size.
In cases where the nuclear quantum effect is non-negligible, more sophisticated approaches like path-integral MD would be needed for properly describing relevant phenomena, which typically require one or two orders of magnitude more computational resources than classical MD.\cite{ko2019isotope,chunyi_jpcb}
Moreover, for systems involving strongly-correlated electronic interactions, methods beyond DFT, such as quantum Monte Carlo,\cite{ceperley1986quantum} will have a more satisfactory accuracy at the price of larger computational cost.
See, e.g., Ref. \citenum{mcmahon2012properties}, for discussions on the properties of hydrogen and helium under extreme conditions and the influence of different simulation methods.


In the last few years, machine-learning (ML) based potentials have been advanced to circumvent the high computational cost of AIMD without loss of accuracy\cite{behler2007generalized,bartok2010gaussian,schutt2017schnet,smith2017ani,zhang2018deep}. 
One of the representatives is the Deep Potential Molecular Dynamics (DeePMD) scheme,\cite{zhang2018deep, zhang2018end} of which the potential energy surface is fitted via a deep neural network to expensive $ab\ initio$ data. 
Studies have demonstrated that for a wide variety of systems,  DeePMD simulations possess accuracy comparable to that of AIMD and efficiency competitive to classical MD simulations.
We refer to Ref.~\citenum{wen2022deep} for a thorough review of some recent development of DeePMD for materials science.
Notwithstanding the successes achieved via ML-based potentials, obstacles still remain in the scenario that requires comprehensive description offered by high-level QM methods.
The training of DeePMD model usually demands thousands of QM-labeled frames, which might become a bottleneck when the QM method of choice is expensive.
As shown in Table S1, for example, for a 64-water-molecule system, within the DFT framework, the computational costs using different functionals can differ by nearly three orders of magnitude.
Indeed, the bridge that efficiently connects time-consuming QM calculations 
and ML-based potential energy models is yet to be assembled so as to alleviate or even eliminate such computational bottleneck. 

Proposed in 2020, the Deep Kohn-Sham (DeePKS) approach introduces a general framework for generating highly accurate self-consistent energy functionals with remarkably reduced computational cost,\cite{chen2020ground,chen2020deepks}  which makes it an ideal ``bridge'' between expensive QM models and DeePMD. 
While the DeePKS model has been comprehensively tested for isolated molecular systems, we implement it here for periodic systems in an open-source software ABACUS\cite{li_abacus,chen_abacus} and demonstrate that \wh{trained by a considerably small number of data, the DeePKS model reproduces the target energies and forces given by strongly constrained and appropriately normed (SCAN) meta-GGA functional\cite{sun2015SCAN} for salt water and hybrid SCAN0 furnctional\cite{hui2016scan} for pure water at \QO{only a few times more expensive} computational cost 
as \QO{compared to} the Perdew-Burke-Ernzerhof (PBE) GGA functional\cite{perdew1996generalized}.}
The trained DeePKS model is applied in SCF calculations to generate labels for the DeePMD \wh{model}, with an estimated two orders of magnitude saving in computational time for labeling. 
The resulting \wh{DeePMD} model is then employed for MD simulations to compute various structural and dynamical properties of pure and salt water. 
Excellent agreement is found between the DeePKS-DeePMD predicted results and the previously reported data from SCAN/SCAN0-based DeePMD simulations, which highlights the reliability of the bridging role played by the DeePKS model. 

It should be noted that a variety of ML-assisted functionals other than DeePKS have also been developed, such as NeuralXC\cite{neural_xc} and OrbNet\cite{qiao2020orbnet}, which share a similar goal as DeePKS, i.e., to lift the accuracy of baseline functionals towards that provided by more accurate methods via a ML-based model while maintaining their efficiency. 
Other ML-based functionals, including DM21\cite{dm21} and SCAN-L\cite{scanl}, are developed to pave the way toward exact universal functional. 
Here we apply DeePKS in this work to demonstrate the capability of such ML-based functional in connection with ML-based potentials.
From a practical point of view, we see this work as a timely contribution to this rapidly developing field, and we stress that with a series of open-source implementations of the methodology, including DeePKS-kit\cite{chen2020deepkskit} for the training and generation of the DeePKS model, ABACUS for DeePKS-based DFT calculations for periodic systems, as well as DeePMD-kit\cite{wang2018kit} and DP-GEN\cite{zhang2020dp} for the training and generation of \wh{DeePMD} models, various applications demanding QM accuracy at a higher level will be made computationally feasible. 

\section{Method}
\label{sec::method}
\subsection{DeePKS for isolated systems}
Before introducing the DeePKS formalism for periodic systems, we briefly review the case of isolated systems.
We consider the many-body Schr\"odinger equation of $N$ electrons:
\begin{equation}
    (\hat{T}+\hat{V}_{ee}+\hat{V}_{\scriptsize{\textrm{ext}}})\Psi_0 = E \Psi_0 ,
\end{equation}
where $\hat{T}$, $\hat{V}_{ee}$, and $\hat{V}_{\scriptsize{\textrm{ext}}}$ are the operators for kinetic, electron-electron interaction, and external potential, respectively, $E$ is the ground-state energy of the system, and $\Psi_0$ represents the ground-state $N$-electron wavefunction.

In the standard Kohn-Sham scheme\cite{kohn1965self, hohenberg1964inhomogeneous}, one employs an auxiliary non-interacting system under an effective external potential $\hat{V}_{\scriptsize{\textrm{KS}}}$, which yields the same ground-state electron density as the original interacting system. 
The auxiliary system can thus be represented by a single Slater determinant of a set of one-particle eigenstates $\{\phi_i\}$, obtained by self-consistently solving the single particle Hamiltonian:
\begin{equation}
    \hat{h}_i \phi_i = \epsilon_i \phi_i ,
\end{equation}
where $\hat{h}_i = \hat{T}+\hat{V}_{\scriptsize{\textrm{KS}}}$.

Conventionally, the effective potential $\hat{V}_{\scriptsize{\textrm{KS}}}$ is partitioned into three components:
\begin{equation}
    \hat{V}_{\scriptsize{\textrm{KS}}} = \hat{V}_{\scriptsize{\textrm{ext}}} + \hat{V}_{\scriptsize{\textrm{H}}} + \hat{V}_{\scriptsize{\textrm{XC}}},
\end{equation}
namely, the external potential of the original interacting systems $\hat{V}_{\scriptsize{\textrm{ext}}}$, the Hartree potential, namely the static Coulomb potential produced by the electron density of the system $\hat{V}_{\scriptsize{\textrm{H}}}$, and the exchange-correlation potential $\hat{V}_{\scriptsize{\textrm{XC}}}$, which captures the remaining electron-electron interactions.

The exact form of the exchange-correlation potential still remains elusive. 
The major task in Kohn-Sham DFT is thus to devise better approximations of the exchange-correlation functional. 
In the traditional Kohn-Sham scheme,\cite{hohenberg1964inhomogeneous,kohn1965self} 
$\hat{V}_{\scriptsize{\textrm{XC}}}$ is spatially local, while in the generalized Kohn-Sham scheme,\cite{becke} $\hat{V}_{\scriptsize{\textrm{XC}}}$ includes non-local contributions. 

The DeePKS scheme seeks a Hamiltonian in the generalized Kohn-Sham framework by connecting a baseline method and a reference method through a neural network model. 
Typically, baseline methods are chosen to be lower level methods that are computationally efficient but lack the desired level of accuracy for the problem under consideration; while the reference methods are high level methods that are accurate but computationally expensive.

The basic idea is to partition the Hamiltonian into two parts:
\begin{equation}
    \hat{h}_{\scriptsize{\textrm{DeePKS}}} = \hat{h}_{\scriptsize{\textrm{baseline}}} + \hat{V}^\delta.
\end{equation}
The first part is the Hamiltonian of the baseline method, while the second part is the correction potential provided by the neural network model. 
As a result, the total energy is also partitioned into two parts:
\begin{equation}
    E_{\scriptsize{\textrm{DeePKS}}} = E_{\scriptsize{\textrm{baseline}}} + E_\delta.
\end{equation}

In this work, we solve the Hamiltonian in the basis of numerical atomic orbitals $\{\chi_\mu\}$,~\cite{chen_abacus, chen2011electronic, blum2009ab, soler2002siesta, Ozaki2003variationally} and the neural network contribution term $V_\delta$ is constructed based on projected density matrices:
\begin{equation}
    D_{nlmm'}^I = \sum_{\mu\nu} \rho_{\mu\nu} \langle \chi_\mu | \alpha_{nlm}^I \rangle \langle \alpha_{nlm'}^I |\chi_\nu \rangle,
\end{equation}
where $\rho_{\mu\nu}$ is the density matrix of the system, and $\{|\alpha\rangle\}$ is a set of localized orbitals centered on atoms, labeled by atomic index $I$, and quantum numbers $nlm$. 

To preserve the rotational invariance, we further take the eigenvalues of blocks of projected density matrices with the same indices $I$, $n$ and $l$ to obtain a series of descriptors:

\begin{equation}
    \mathbf{d}^I_{nlm} = Eig(D_{nlmm'}^I).
\end{equation}

In some cases, this eigenvalue decomposition step introduces discontinuities due to the sorting of eigenvalues, and may cause convergence problems when applying the model in SCF calculations. 
To circumvent this issue, there is an option to further symmetrize the descriptors:

\begin{equation}
    \mathbf{d}^{I,symm}_{nlm} = g_{\scriptsize{\textrm{symm}}}(\mathbf{d}^I_{nlm}).
\end{equation}
where $g_{\scriptsize{\textrm{symm}}}$ is a symmetrization function which is invariant under the permutation of its arguments.
In our current implementation, $g_{\scriptsize{\textrm{symm}}}$ is chosen to be thermal averaging, and details of the symmetrization step as well as the neural network structure can be found in the Appendix of Ref.~\citenum{chen2020deepkskit}.

The descriptors are then grouped into vectors according to the atomic index $I$, and the correction energy term becomes a summation of atomic contributions:
\begin{equation}
    E_\delta = \sum_{I} \mathrm{F}_{\scriptsize{\textrm{NN}}}(\mathbf{d}^I|\mathbf{\omega}),
\end{equation}
where $\mathbf{\omega}$ is the vector of parameters for the deep neural network $\mathrm{F}_{\scriptsize{\textrm{NN}}}$. By calculating a set of reference systems, we have the target energies $E_{\scriptsize{\textrm{target}}}$, the baseline energies $E_{\scriptsize{\textrm{baseline}}}$, 
as well as the descriptors generated by the baseline method $\mathbf{d}^I$. The training of $\mathrm{F}_{\scriptsize{\textrm{NN}}}$ is then carried out by using the energy difference $E_{\scriptsize{\textrm{target}}}-E_{\scriptsize{\textrm{baseline}}}$ as the label.


With the expression for $E_\delta$, the corresponding matrix elements of the correction potential are given by:
\begin{align}
\hat{V}^\delta_{\mu\nu} = & \frac{\partial E_\delta}{\partial \rho_{\mu\nu}} \nonumber \\
                 = & \sum_{Inlmm'} \frac{\partial E_\delta}{\partial D_{nlmm'}^I} \frac{\partial D_{nlmm'}^I}{\partial \rho_{\mu\nu}} \nonumber \\
                = & \sum_{Inlmm'} \frac{\partial E_\delta}{\partial D_{nlmm'}^I} \langle \chi_\mu | \alpha_{nlm}^I \rangle \langle \alpha_{nlm'}^I |\chi_\nu \rangle.
\end{align}

Solving the Hamiltonian $\hat{h}_{\scriptsize{\textrm{DeePKS}}} = \hat{h}_{\scriptsize{\textrm{baseline}}} + \hat{V}^\delta$ gives a set of ground state wavefunctions $\{\phi_i|\omega\}$ and ground state energy $E_{\scriptsize{\textrm{DeePKS}}} = E_{\scriptsize{\textrm{baseline}}}[\{\phi_i|\mathbf{\omega}\}] + E^\delta[\{\phi_i|\mathbf{\omega}\},\mathbf{\omega}]$.
However, in general there is a discrepancy between the $E_{\textrm{DeePKS}}$ here and the target energy $E_{\scriptsize{\textrm{target}}}$. The origin of this discrepancy comes from the fact that the ground state of $\hat{h}_{\scriptsize{\textrm{DeePKS}}}$ is different from that of the initial baseline method $\hat{h}_{\scriptsize{\textrm{baseline}}}$.

As a result, the training of DeePKS adopts an iterative strategy, where the vector of model parameters $\mathbf{\omega}$ is updated through training, followed by solving the new $\hat{h}_{\scriptsize{\textrm{DeePKS}}}$ to get a new set of descriptors and labels. The process is repeated until convergence is achieved.

For later iterations, we can also calculate the total force under $\hat{h}_{\scriptsize{\textrm{DeePKS}}}$, given by:
\begin{align}
    \mathbf{F}_{\scriptsize{\textrm{DeePKS}}}[\{\phi_i|\omega\}] = & \mathbf{F}_{\scriptsize{\textrm{baseline}}}[\{\phi_i|\omega\}] - \frac{\partial E^\delta [\{\phi_i|\omega\}]}{\partial \mathbf{X}} \nonumber\\
        = & \mathbf{F}_{\scriptsize{\textrm{baseline}}}[\{\phi_i|\omega\}] - \sum_{Inlmm'} \frac{\partial E_\delta}{\partial D_{nlmm'}^I} \sum_i \frac{d}{d\mathbf{X}}[f_i\langle \phi_i | \alpha_{nlm}^I \rangle \langle \alpha_{nlm'}^I |\phi_i \rangle].
\end{align}
where $f_i$ is the occupation number of orbital $\phi_i$.
As our goal is to reproduce the total energies and forces of the target method, we also include force term in the loss function $L(\omega)$, and the optimization problem now becomes:
\begin{equation}\label{eqn::target}
\min_\omega L(\omega), ~    
L(\omega)=|E_{\scriptsize{\textrm{target}}}-E_{\scriptsize{\textrm{DeePKS}}}[\{\phi_i|\omega\}]|^2 + \lambda |\mathbf{F}_{\scriptsize{\textrm{target}}}-\mathbf{F}_{\scriptsize{\textrm{DeePKS}}}[\{\phi_i|\omega\}]|^2,
\end{equation}
where the weighting factor $\lambda$ is adjusted in the iterative training process to balance the error in energy and force. Its value typically falls in the range of 1 to 50. Here the notation $\{\phi_i|\mathbf{\omega}\}$ is used to emphasize that the eigenstates ${\phi_i}$ of the DeePKS Hamiltonian $\hat{h}_{\scriptsize{\textrm{DeePKS}}} = \hat{h}_{\scriptsize{\textrm{baseline}}} + \hat{V}^\delta$ depend on the expression of $\hat{V}^\delta$, hence on the neural network parameters $\mathbf{\omega}$.

We refer to Ref.~\citenum{chen2020deepks} and Ref.~\citenum{chen2020deepkskit} for more details of the DeePKS formalism and the training strategy, including the treatment of force labels, as well as the construction of the neural network.

\subsection{DeePKS for Periodic Systems}
For periodic systems, the external potential possesses the translational symmetry:
\begin{equation}
    \hat{V}_{\scriptsize{\textrm{ext}}}(\mathbf{r}-\mathbf{R}) = \hat{V}_{\scriptsize{\textrm{ext}}}(\mathbf{r}),
\end{equation}
where $\mathbf{R}$ is the lattice vector used to label the unit cells in the periodic lattice.
According to the Bloch theorem, the Kohn Sham eigenstates of the system are expressed as:
\begin{equation}
    \phi_{i\mathbf{k}}(\mathbf{r}) = e^{\mathrm{i}\mathbf{k}\mathbf{r}}u_{i\mathbf{k}}(\mathbf{r}),
\end{equation}
where $\mathbf{k}$ is the reciprocal space lattice vector and the quantum number $i$ labels the band index.
%
The Bloch wavefunction $u_{i\mathbf{k}}(\mathbf{r})$ has the same periodicity of the external potential $\hat{V}_{\scriptsize{\textrm{ext}}}$ and can be solved by diagonalization of the Hamiltonian $H(\mathbf{k})$.

To obtain the matrix elements of $H(\mathbf{k})$ in the atomic basis $\{\chi_\mu\}$, we first calculate:
\begin{equation}
    H_{\mu\nu}(\mathbf{R}) = \langle \chi_{\mu \mathbf{R}}|\hat{h}|\chi_{\nu \mathbf{0}}\rangle, 
\end{equation}
where $\chi_{\mu \mathbf{R}}$ is the periodic image of atomic basis $\chi_{\mu}$ in the unit cell $\mathbf{R}$, namely $\chi_{\mu \mathbf{R}}(\mathbf{r}) = \chi_\mu(\mathbf{r}-\mathbf{R})$.

The Hamiltonian for single $\mathbf{k}$-point $H(\mathbf{k})$ is then given by:
\begin{equation}
    H(\mathbf{k}) = \sum_\mathbf{R} e^{-\mathrm{i}\mathbf{k}\mathbf{R}} H(\mathbf{R}).
\end{equation}

All physical quantities are obtained as an average over single $k$ points.
For example, the electron density is given by:
\begin{align}
    \rho (\mathbf{r}) = &\frac{1}{N_\mathbf{k}} \sum_{i\mathbf{k}} f_{i\mathbf{k}} \phi_{i\mathbf{k}}^*(\mathbf{r}) \phi_{i\mathbf{k}}(\mathbf{r}) \nonumber\\
             = &\frac{1}{N_\mathbf{k}} \sum_{\mu\nu} \sum_\mathbf{R} \sum_{\mathbf{k}} \rho_{\mu \nu}(\mathbf{k}) \chi_{\mu \mathbf{R}}^*(\mathbf{r}) \chi_{\nu\mathbf{0}}(r) e^{-\mathrm{i}\mathbf{k}\mathbf{R}} \nonumber\\
             = &\sum_{\mu\nu}\sum_\mathbf{R} \rho_{\mu\nu}(\mathbf{R}) \chi_{\mu \mathbf{R}}^*(\mathbf{r}) \chi_{\nu\mathbf{0}}(\mathbf{r}),
\end{align}
where we define the real-space density matrix as:
\begin{equation}
    \rho_{\mu\nu}(\mathbf{R}) = \frac{1}{N_\mathbf{k}} \sum_{\mathbf{k}} \rho_{\mu \nu}(\mathbf{k}) e^{-\mathrm{i}\mathbf{k}\mathbf{R}}.
\end{equation}

Similarly, the projected density matrix used to construct descriptors is calculated as:
\begin{align}
    D_{nlmm'}^I & = \frac{1}{N_\mathbf{k}} \sum_{\mu\nu} \sum_\mathbf{k} \sum_{\mathbf{R}'} \rho_{\mu\nu}(\mathbf{k}) \langle \chi_{\mu \mathbf{R}} | \alpha_{nlm\mathbf{R}'}^I \rangle \langle \alpha_{nlm'\mathbf{R}'}^I |\chi_{\nu\mathbf{0}} \rangle e^{-\mathrm{i}\mathbf{k}\mathbf{R}} \nonumber\\
                & = \sum_{\mathbf{R}\mathbf{R}'} \sum_{\mu\nu} \rho_{\mu\nu}(\mathbf{R}) \langle \chi_{\mu \mathbf{R}} | \alpha_{nlm\mathbf{R}'}^I \rangle \langle \alpha_{nlm'\mathbf{R}'}^I |\chi_{\nu\mathbf{0}} \rangle.
\end{align}

Then, the contribution of $\hat{V}_\delta$ to the real-space Hamiltonian is derived as:
\begin{align}
    \hat{V}^\delta_{\mu\nu}(\mathbf{R}) = & \sum_{Inlmm'} \frac{\partial E_\delta}{\partial D_{nlmm'}^I} \frac{\partial D_{nlmm'}^I}{\partial \rho_{\mu\nu}(\mathbf{R})} \\
        = & \sum_{Inlmm'} \sum_{\mathbf{R}'} \frac{\partial E_\delta}{\partial D_{nlmm'}^I} \langle \chi_{\mu \mathbf{R}} | \alpha_{nlm\mathbf{R}'}^I \rangle \langle \alpha_{nlm'\mathbf{R}'}^I |\chi_{\nu\mathbf{0}} \rangle.
\end{align}

As noted in the previous section, we also include force label in the training process, with target function given in Eq. \ref{eqn::target}. The total force in the case of multiple $k$ points is given by:
\begin{align}
    \mathbf{F}_{\scriptsize{\textrm{DeePKS}}}[\{\phi_i|\omega\}] = & \mathbf{F}_{\scriptsize{\textrm{baseline}}}[\{\phi_i|\omega\}] - \frac{\partial E^\delta [\{\phi_i|\omega\}]}{\partial \mathbf{X}} \nonumber\\
        = & \mathbf{F}_{\scriptsize{\textrm{baseline}}}[\{\phi_i|\omega\}] - \sum_{Inlmm'} \frac{\partial E_\delta}{\partial D_{nlmm'}^I} \sum_{i\mathbf{k}} \frac{d}{d\mathbf{X}}[f_{i\mathbf{k}}\langle \phi_{i\mathbf{k}} | \alpha_{nlm}^I \rangle \langle \alpha_{nlm'}^I |\phi_{i\mathbf{k}} \rangle] \nonumber\\
        = & \mathbf{F}_{\scriptsize{\textrm{baseline}}}[\{\phi_i|\omega\}] - \sum_{Inlmm'} \frac{\partial E_\delta}{\partial D_{nlmm'}^I} \sum_{\mathbf{RR}'} \rho_{\mu\nu}(\mathbf{R}) \frac{d}{d\mathbf{X}}[\langle \chi_{\mu \mathbf{R}} | \alpha_{nlm\mathbf{R}'}^I \rangle \langle \alpha_{nlm'\mathbf{R}'}^I |\chi_{\nu \mathbf{0}}\rangle].
\end{align}

\subsection{Computational Details}
\begin{figure}[htp]
    \includegraphics[width=0.5\linewidth]{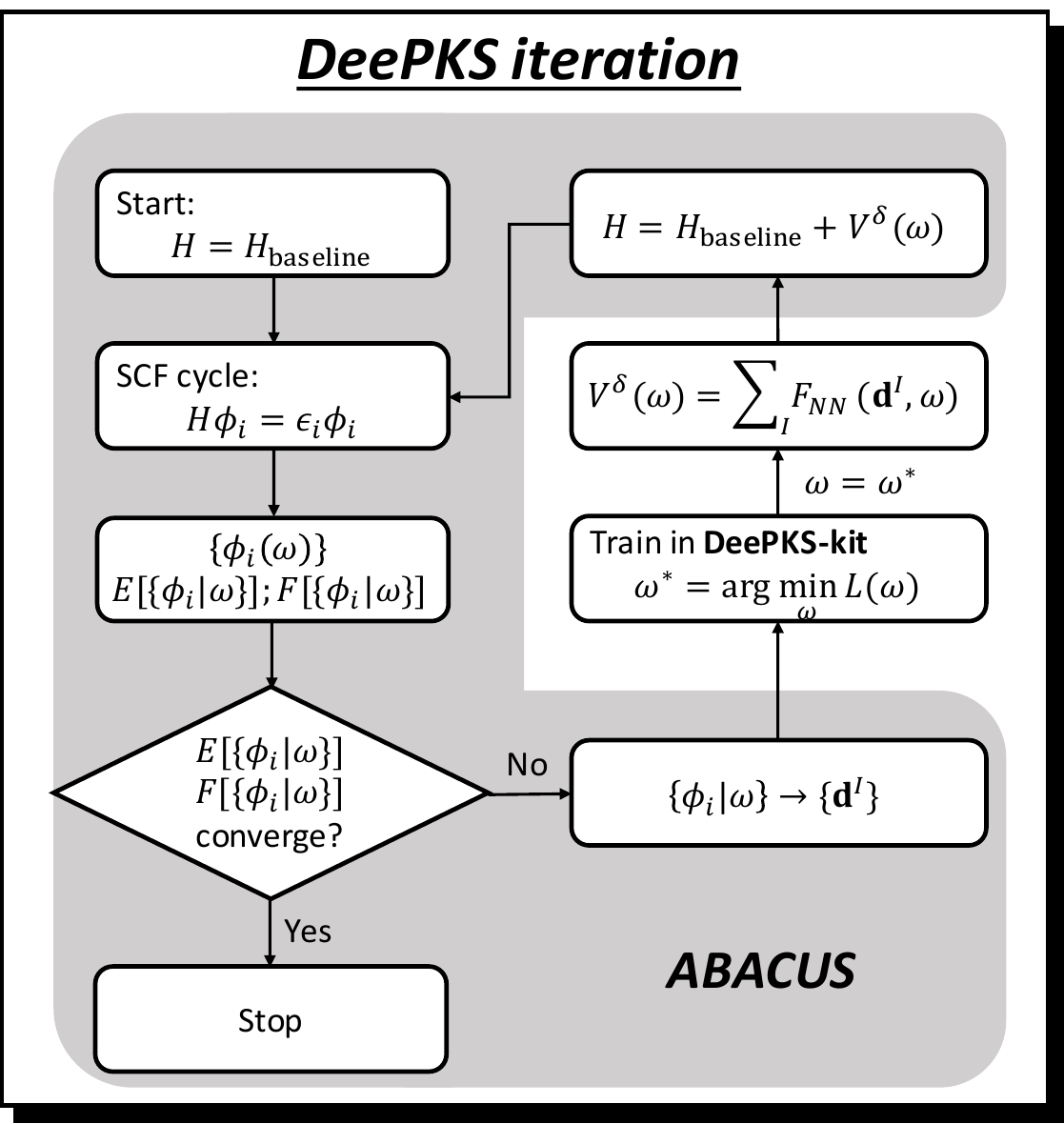}
    \caption{Flowchart of the DeePKS iterative training implemented within the ABACUS density functional theory package. 
    }
    \label{fig::flowchart}
\end{figure}

As mentioned in previous sections, the training of DeePKS adopts an iterative strategy, which alternates between training the neural network $\mathrm{F}_{\scriptsize{\textrm{NN}}}(\mathbf{d}^I|\mathbf{\omega})$ and solving SCF with $\hat{H}_{\scriptsize{\textrm{DeePKS}}}=\hat{H}_{\scriptsize{\textrm{baseline}}}+\hat{V}^\delta$. 
While the training step is performed within DeePKS-kit,\cite{chen2020deepkskit} the package \textit{per se} does not contain the functionality of solving SCF. Instead, an existing SCF software is invoked for such purpose. 

We have implemented the DeePKS method in the ABACUS\cite{chen_abacus,li_abacus} package, which supports both numerical atomic orbitals and plane-wave basis with the periodic boundary conditions. The ABACUS package can be freely downloaded online~\cite{abacus_website}.
For the numerical atomic orbitals that form the projectors, the radial parts of $\{|\alpha\rangle\}$ are chosen to be the spherical bessel functions, namely:
\begin{equation}
    \alpha_{nlm}(\mathbf{r}) = f_{nl}(r)Y_{lm}(\theta,\phi),
\end{equation}
where
\begin{equation}
    f_{nl}(r) = \begin{cases}
        j_{nl}(q_nr) & (r\leq r_c) \\
        0 & (otherwise).
\end{cases}
\end{equation}
Here $r_c$ is the radius cutoff, and $q_n$ is chosen to ensure $j_{nl}(q_nr_c)=0$. A kinetic energy cutoff is imposed to determine the upper bound for the value of $q_n$, hence the number of spherical bessel functions. Typically, the kinetic energy cutoff is set to be the same as the underlying SCF calculations. 

In this work, we used a radius cutoff of 5 Bohr, and the kinetic energy cutoff is set to be 100 Ry, with $l=0,1,2$. This resulted in a total number of 15 spherical bessel functions per $l$ channel, giving an overall of 135 descriptors per atom.
More details on the spherical bessel functions and the evaluation of orbital overlaps $\langle \alpha_{nlm}|\chi_\mu\rangle$ can be found in Ref. \citenum{chen_abacus} and Ref. \citenum{li_abacus}.
%

The implemented DeePKS iterative training process is summarized in Fig.~\ref{fig::flowchart}. 
The steps in the grey region are those carried out by ABACUS. In each iteration, ABACUS reads the neural network model file provided by DeePKS-kit \MC{and} calculates the desired matrix elements $\hat{V}^\delta_{\mu\nu}$, then solves the DeePKS Hamiltonian $\hat{H}_{\scriptsize{\textrm{DeePKS}}}$ and outputs the descriptors and labels in the format that is readable by DeePKS-kit.



We notice that a practical challenge for testing the performance of the DeePKS+ABACUS scheme is that we need extensively generated high-level electronic structure data for benchmark purposes. 
As such, we chose two representative datasets that have already been well benchmarked in recent works and used to train \QO{DeePMD potential} 
models for important applications.
The first dataset\cite{chunyi_jpcb} contains water snapshots from both classical and Feynman path-integral molecular dynamics calculations with energy and force labels at the SCAN0 level.
A Deep Potential model was generated from this dataset and used to calculate several properties of water, and later used to investigate the many-body effects in the X-ray absorption spectra of liquid water\cite{Tang2022many}.
The second dataset\cite{chunyi_nc} was generated via a concurrent learning approach\cite{zhang2019active} and used to train Deep Potentials for modeling the structural properties of sodium chloride solutions at different concentrations at the level of the SCAN functional. Additional computational details for these two datasets can be found in the Supporting Information.

We used the SCAN0 AIMD trajectories of 64 water molecules to compare the sample efficiency of the DeePKS model and the DeePMD model.
Next, for both datasets, we chose the PBE functional as the baseline model and used only a small group of samples to obtain reliable DeePKS models at production level.
We tested the validity of the resultant DeePKS models by relabeling a much larger group of samples from the same datasets with DeePKS and carried out DeePMD training. 
The DeePMD models were in turn applied to run MD simulations using LAMMPS\cite{Plimpton1995}. 
Structural and thermodynamic properties, including the radial distribution function (RDF), bulk density, and others, were calculated and compared with existing results.

\section{Result and Discussion}
\label{sec::result}
\subsection{DeePKS learning curves with respect to training samples}
\begin{figure*}[htp]
    \includegraphics[width=0.5\linewidth]{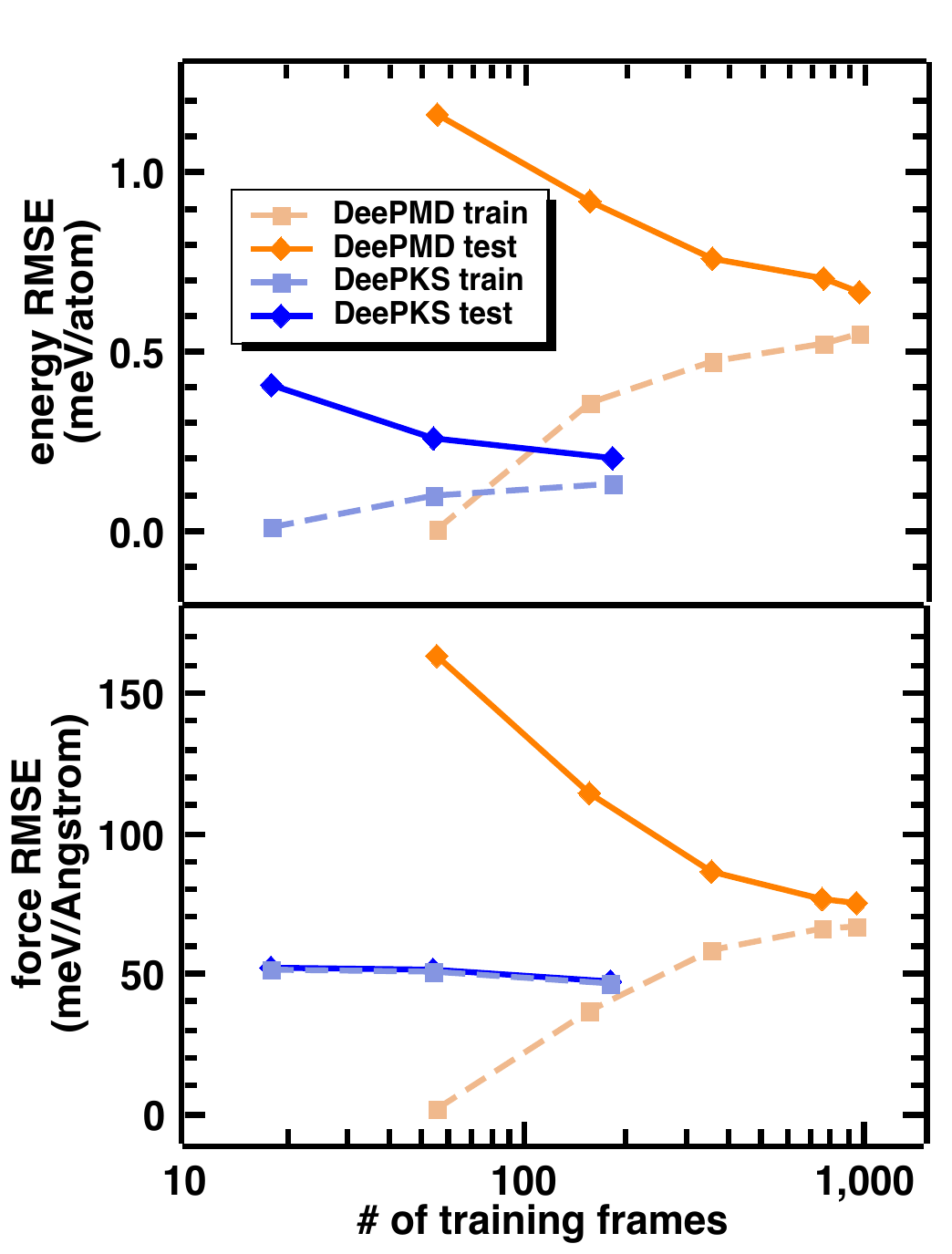}
    \caption{Learning curves for energy (upper panel) and force (lower panel) given by DeePMD (orange) and DeePKS (blue) with respect to the number of training frames. Dashed line with squares indicates train set error; solid line with diamonds indicates test set error.}
    \label{fig::learning_curve}
\end{figure*}
To explore the capability and generalizability of the DeePKS model, we construct three training sets by \wh{randomly (subject to uniform distribution)} picking 18, 54, and 180 frames from the previously reported SCAN0 AIMD trajectories of 64 water molecules,~\cite{chunyi_jpcb} and train the PBE-based DeePKS model with SCAN0 energy and force labels.
Similar uniform sampling from these SCAN0 trajectories, with larger sampling sizes (55, 155, 355, 755, and 955 frames), is applied to the training of the DeePMD model so as to make a comparison between these two models. The learning curves of DeePMD and DeePKS with respect to the size of the training set are given in Fig.~\ref{fig::learning_curve}. 
It can be seen that with significantly fewer frames, the DeePKS model provides more accurate predictions as compared to DeePMD model. The generalization gap of DeePKS model is also notably smaller than that of DeePMD model. 
It is shown in Table S1 that the SCAN0 SCF result for 64 water molecules, which takes more than a day to be obtained, can be accurately reproduced within a quarter of an hour by applying the DeePKS model, which corresponds to more than two orders of magnitude savings in time.
The take-home message conveyed by Fig. \ref{fig::learning_curve} and Table S1 is that \QO{the training process of the DeePMD potential, which originally demands more than a thousand expensive SCAN0 jobs, can be effectuated with around one hundred SCAN0 jobs plus a thousand significantly faster DeePKS jobs. In other words}, the DeePKS model can serve as a bridge that connects the expensive $ab\ initio$ calculations such as SCAN0 DFT and the machine learning potentials, and remarkably reduces the effort required in the MD simulations at higher rung of the Jacob's ladder.

\subsection{Modelling liquid water}
For systems consisting of 64 water molecules, we perform SCF calculations on 1022 unique structures randomly (subject to uniform distribution) picked from previous DeePMD (with SCAN0 label) and SCAN0 AIMD modelling results (from Ref. \citenum{chunyi_jpcb}) with the DeePKS model trained via 180 training samples. 1000 out of 1022 SCF calculations reach the convergence threshold, and these 1000 converged energies and forces are applied as labels for DeePMD training. 
The resulting Deep Potential model is then employed in LAMMPS for molecular dynamics simulation of 512 water molecules.
Various structural properties and the diffusion coefficient are explored and compared with previously reported results.
All structural properties are obtained via 30 ps $NpT$ ensemble simulations at 1 bar and 330 K with a time step of 0.5 fs with the first 10 ps discarded for equilibrium, while the diffusion coefficient is obtained via 300 ps $NVE$ ensemble simulations with the cell size fixed at the value obtained from the $NpT$ simulation. 

\begin{figure}[htp]
    \includegraphics[width=1.0\linewidth]{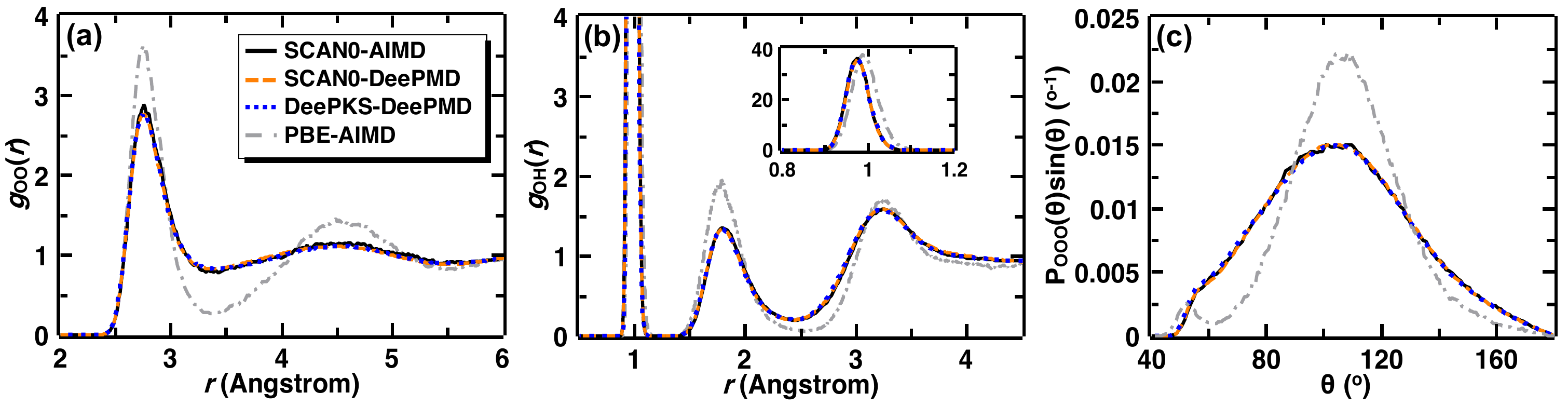}
    \caption{RDFs {\bf(a)} $g_{\scriptsize{\textrm{OO}}}(r)$, {\bf(b)} $g_{\scriptsize{\textrm{OH}}}(r)$, and {\bf{(c)}} bond angle distribution $P_{\scriptsize{\textrm{OOO}}}(\theta)$ given by DeePKS-DeePMD (blue dotted line), SCAN0-AIMD (black solid line) from Ref. \citenum{chunyi_jpcb}, SCAN0-DeePMD (orange dashed line) from Ref. \citenum{chunyi_jpcb}, and PBE-AIMD (gray dotted-dashed line) from Ref. \citenum{mohan_pnas}.}
    \label{fig::water_scan0}
\end{figure}
We first analyze the RDFs, which correspond to the probability of finding a given pair of atoms as a function of distance in real space, with DeePKS-DeePMD simulations. The resulting oxygen-oxygen and oxygen-hydrogen RDFs, $g_{\scriptsize{\textrm{OO}}}(r)$ and $g_{\scriptsize{\textrm{OH}}}(r)$ are shown in Fig. \ref{fig::water_scan0}(a) and (b), respectively. the previously computed results via SCAN0-AIMD, SCAN0-DeePMD, and PBE-AIMD are also shown for comparison. 
It can be seen from Figs. \ref{fig::water_scan0}(a) and (b) that the RDFs given by DeePKS-DeePMD simulations are in good agreement with both the SCAN0-AIMD and the SCAN0-DeePMD results, including the significantly less overstructured peaks of $g_{\scriptsize{\textrm{OO}}}(r)$ and the slightly shortened O-H covalent bond length (indicated by the first peak of $g_{\scriptsize{\textrm{OH}}}(r)$) as compared to the PBE result. Similar observations are found in the bond angle distribution ($P_{\scriptsize{\textrm{OOO}}}(\theta)$) analysis as shown in Fig. \ref{fig::water_scan0}(c), which quantifies the three-body correlations in water. While PBE predicts a narrower bond angle distribution, DeePKS is able to quantitatively reproduce the distribution predicted by SCAN0. Overall, for liquid water, the overstructuring issue in PBE functional is remarkably alleviated with the trained DeePKS model, which provides almost identical structural properties as compared to SCAN0 results. 
 
 Next, we explore the bulk density and H-bonds of liquid water with DeePKS-DeePMD simulations. Note that the diffusion coefficient is computed for both water and deuterated water so as to provide a more comprehensive comparison with the SCAN0 results from Ref. \citenum{chunyi_jpcb}. By including the description of nondirectional van der Waals (vdW) interaction on intermediate length-scales, SCAN0 predicts a more disordered and compact water structure, leading to a higher bulk density and weakened H-bond strength. As shown in Table \ref{table::water_scan0}, the bulk density predicted via DeePKS-DeePMD (1.024 g/cm$^3$) is consistent with that predicted by SCAN0-DeePMD (1.030 g/cm$^3$), and is notably larger than that predicted via PBE-AIMD (0.850 g/cm$^3$). The average number of H-bonds per water molecule computed with DeePKS-DeePMD is 3.58, which is identical to that given by SCAN-DeePMD and notably smaller than the PBE-AIMD result (3.77 according to Ref. \citenum{mohan_pnas}). The weakened H-bond strength is also evidenced by the more dominant region between the first peak and the second peak of $g_{\scriptsize{\textrm{OO}}}(r)$ predicted by SCAN0 (as shown in Fig. \ref{fig::water_scan0}(a)), which mainly comprises non-H-bonded molecules that occupy interstitial space between H-bonded ones. 
 
The dynamic property of liquid water we examine in this work with DeePKS-DeePMD is the diffusion coefficient, for both normal and deuterated water. The diffusion of liquid water depends on the formation and breakage of H-bonds through thermal fluctuations. Weakened H-bond strength predicted by SCAN0 and DeePKS (as illustrated above) escalates the tendency of H-bond-breaking and consequently increases the diffusion coefficient. 
It can be seen in Table \ref{table::water_scan0} that $D$ predicted by DeePKS-DeePMD is in excellent agreement with the one predicted by SCAN0-DeePMD, which is one order of magnitude larger than the PBE-AIMD result. 
The same consistency is also observed for the case of deuterated water. 
The good agreement on these structural and dynamical properties highlight the fact that the intermediate-ranged vdW interactions in liquid water, which are intrinsically missed in PBE functional, are successfully captured via the 
\wh{trained DeePKS model using the PBE functional as its baseline,}
and properties of liquid water with expensive \wh{hybrid} meta-GGA (SCAN0) quality can now be much more efficiently predicted within the time comparable to a few PBE jobs.  
 
 \begin{table}[htp]
\tabcolsep 12pt
\centering
\caption{Bulk density ($\rho$), average number of H-bonds per water molecule ($N_{\scriptsize{\textrm{HB}}}$), and diffusion coefficients ($D$) predicted by \MC{DeePKS-DeePMD, SCAN0-DeePMD, and PBE-AIMD} simulations at 330K with 512 water molecules.\textsuperscript{\textit{a}} }
\begin{tabular}{ccccc}
\hline\hline
Method & $\rho$/g$\cdot$cm$^{-3}$ & $N_{\scriptsize{\textrm{HB}}}$ &  $D$/\AA$\cdot$ps$^{-1}$ & deuterated $D$/\AA$\cdot$ps$^{-1}$ \\
\hline
\MC{DeePKS-DeePMD} & 1.024$\pm$0.010 & 3.58 & 0.254$\pm$0.024 & 0.234$\pm$0.019 \\
\MC{SCAN0-DeePMD}\cite{chunyi_jpcb} & 1.030 & 3.58 & 0.251 & 0.223\\
\MC{PBE-AIMD}\cite{mohan_pnas} & 0.850$\pm$0.016 & 3.77 & 0.018$\pm$0.002  & NA\\
exp\textsuperscript{\textit{b}}  & 0.997\cite{nist} & 3.58\cite{hbonds} & 0.24\cite{water_diffusion} & 0.20\cite{water_diffusion}\\
\hline\hline
\end{tabular}

\textsuperscript{\textit{a}}All error bars correspond to one standard deviation. \\
\textsuperscript{\textit{b}}Experimental values are measured at T=300K. 
\label{table::water_scan0}
\end{table}
  
\subsection{Modelling salt water and high-pressure water}
While the investigation of liquid water with DeePKS DeePMD simulations demonstrates accurate computational results compared to SCAN0 counterparts, modelling the electrolyte structure is rather cumbersome due to the sparsity of ions that requires significantly longer simulation time
 for statistical convergence.
Here, we train a PBE-based DeePKS model for salt water with previously conducted SCAN SCF calculations on various concentrations of NaCl solution as labels. The composition of the DeePKS training set can be found in Table S2. 
SCF calculations with such DeePKS model are then carried out on 5676 frames (varying in concentration as shown in Tabel S2), of which 5406 converged results (including energy and force) are utilized as labels for DeePMD training. With the trained potential, we investigate structural properties and densities for salt water with different concentrations as well as pure water under different pressure via DeePKS-DeePMD in LAMMPS. 

\subsubsection{Comparison with SCAN-AIMD simulations}
\begin{figure*}[htp]
    \includegraphics[width=1.0\linewidth]{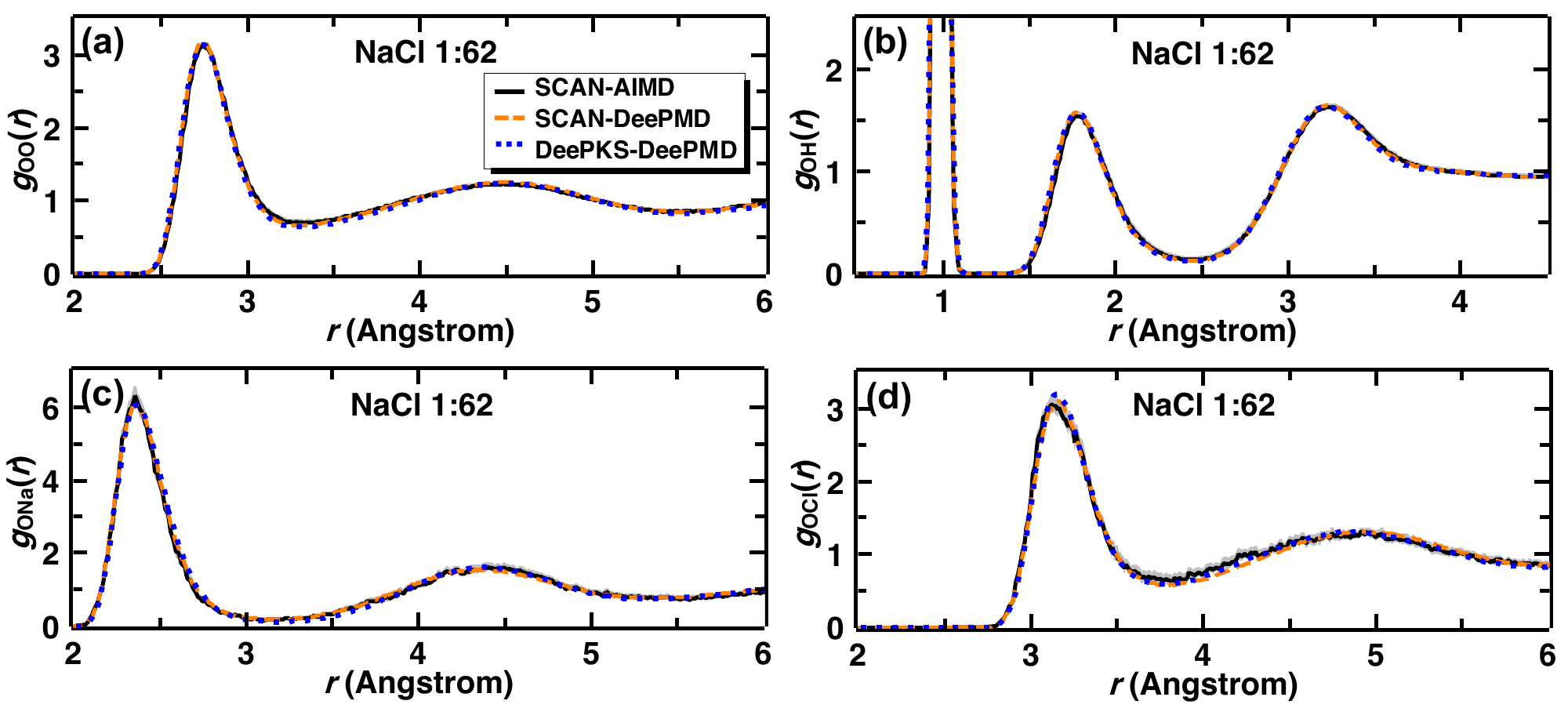}
    \caption{RDFs {\bf(a)} $g_{\scriptsize{\textrm{OO}}}(r)$, {\bf(b)} $g_{\scriptsize{\textrm{OH}}}(r)$, {\bf{(c)}} $g_{\scriptsize{\textrm{ONa}}}(r)$, and {\bf{(d)}} $g_{\scriptsize{\textrm{OCl}}}(r)$ given by DeePKS-DeePMD (blue dotted line), and SCAN-AIMD (black solid line) as well as SCAN-DeePMD (orange dashed line) from Ref. \citenum{chunyi_nc}. Gray shaded area corresponds to one standard deviation from DeePKS-DeePMD statistics with 100 ps time interval.}
    \label{fig::rdf_1_62_nvt}
\end{figure*}
We first compare the RDFs of 1:62 NaCl solution with previously reported SCAN-AIMD and SCAN-DeePMD results. 
A 2-ns DeePKS-DeePMD simulation is carried out with the $NVT$ ensemble using one cubic cell, which consists of one NaCl ion pair and 62 water molecules at 300 K; 
the setup is consistent with the SCAN-DeePMD simulation conditions in Ref. \citenum{chunyi_nc}.  
Fig. \ref{fig::rdf_1_62_nvt} exhibits four RDFs  $g_{\scriptsize{\textrm{OO}}}(r)$, $g_{\scriptsize{\textrm{OH}}}(r)$, $g_{\scriptsize{\textrm{ONa}}}(r)$, and $g_{\scriptsize{\textrm{OCl}}}(r)$, predicted by DeePKS-DeePMD, SCAN-AIMD, and SCAN-DeePMD. 
Note that the SCAN-AIMD simulation only runs for 100 ps due to the highly time-consuming SCF calculations with the SCAN functional. 
We therefore compute the statistical deviation with 100 ps time interval based on DeePKS-DeePMD trajectories (indicated by the gray shaded area in Fig. \ref{fig::rdf_1_62_nvt}). 
All RDFs predicted by DeePKS-DeePMD simulations are in accordance with the SCAN-DeePMD results and statistically matches with the SCAN-AIMD results, which conceptually proves the reliability of our trained DeePKS model.   

\subsubsection{Comparison with SCAN-DeePMD simulations for high-pressure water}

\begin{figure*}[htp]
    \includegraphics[width=0.5\linewidth]{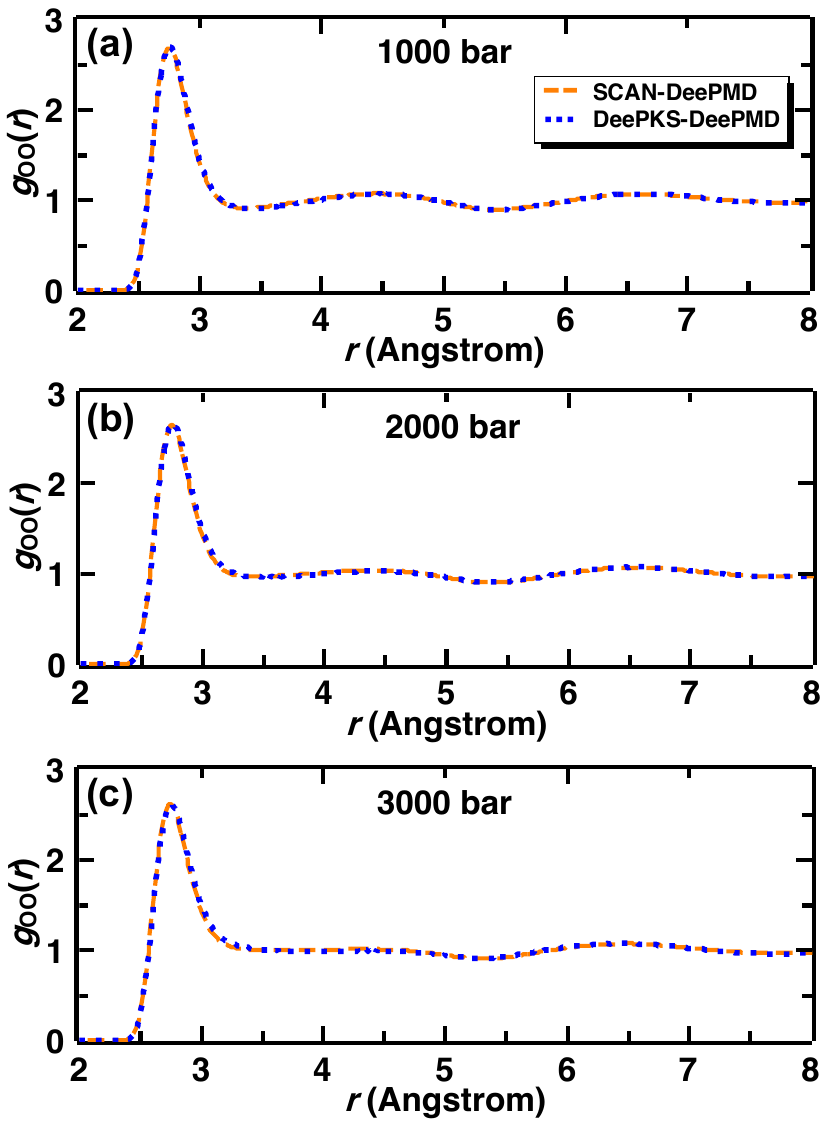}
    \caption{Oxygen-oxygen RDF [$g_{\scriptsize{\textrm{OO}}}(r)$] for pure water at pressure equals {\bf(a)} 1000 bar, {\bf(b)} 2000 bar, and {\bf{(c)}} 3000 bar given by DeePKS-DeePMD (blue dotted line) from this work and SCAN-DeePMD (orange dashed line) from Ref. \citenum{chunyi_nc}.}
    \label{fig::goo_pressure}
\end{figure*}
\begin{figure*}[htp]
    \includegraphics[width=0.4\linewidth]{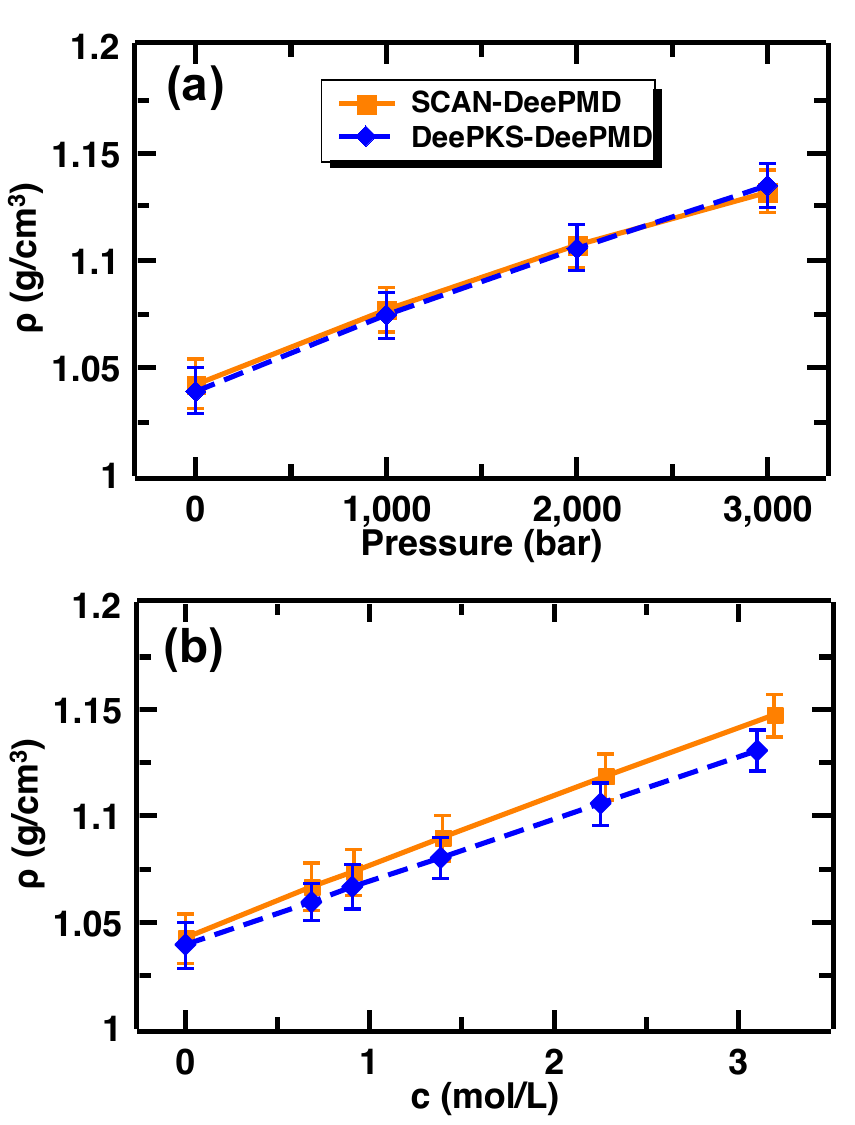}
    \caption{Bulk densities for {\bf{(a)}} pure water at different pressures and {\bf{(b)}} pure water and NaCl solutions with various concentrations at 1 bar predicted by DeePKS-DeePMD (blue diamond) from this work and SCAN-DeePMD (orange square) from Ref. \citenum{chunyi_nc}. The simulation temperature is 330K. Error bars correspond to one standard deviation.}
    \label{fig::density}
\end{figure*}

Structural differences between the high-pressure water and salt water have been comprehensively illustrated in Ref. \citenum{chunyi_nc}. 
Here, with the aforementioned trained Deep Potential based on DeePKS energies and forces, we first investigate the structural properties of water under high pressure via DeePMD simulations using the $NpT$ ensemble with 512 water molecules for 2 ns at 333 K and four different pressures (1 bar, 1 kbar, 2 kbar, and 3 kbar). 
The integration time step is 0.5 fs, with the first 100 ps of trajectory discarded for equilibrium. 
As shown in Fig. \ref{fig::goo_pressure}, the oxygen-oxygen RDFs predicted SCAN-DeePMD under all three high pressures are accurately reproduced by DeePKS-DeePMD simulations. (The comparison at 1 bar will be shown in the following part.) 
It can be seen in Figure S1(a) that as pressure increases, the second and the third coordination shells move inwards, leading to a more compact structure. 
The diminishing feature of the second shell at 3000 bar is consistent with the fact that the tetrahedral network inside liquid water is significantly distorted under high pressure. 
Bulk densities of liquid water have also been calculated as shown in Fig. \ref{fig::density}. Excellent agreement is again observed between SCAN-DeePMD and DeePKS-DeePMD simulations.  

\subsubsection{Comparison with SCAN-DeePMD simulations for salt water with various concentrations}

In this part, we examine our DeePKS model by comparing the structural properties given by DeePKS-DeePMD with those predicted by SCAN-DeePMD for pure and salt water with various concentrations. 
Simulation conditions for DeePKS-DeePMD are kept the same as last section and the pressure is fixed at 1 bar. 
The numbers of NaCl ion pairs and water molecules contained in the periodic cubic cell for DeePMD simulations for each investigated concentration are listed in Table S3. 

\begin{figure*}[htp]
    \includegraphics[width=1.0\linewidth]{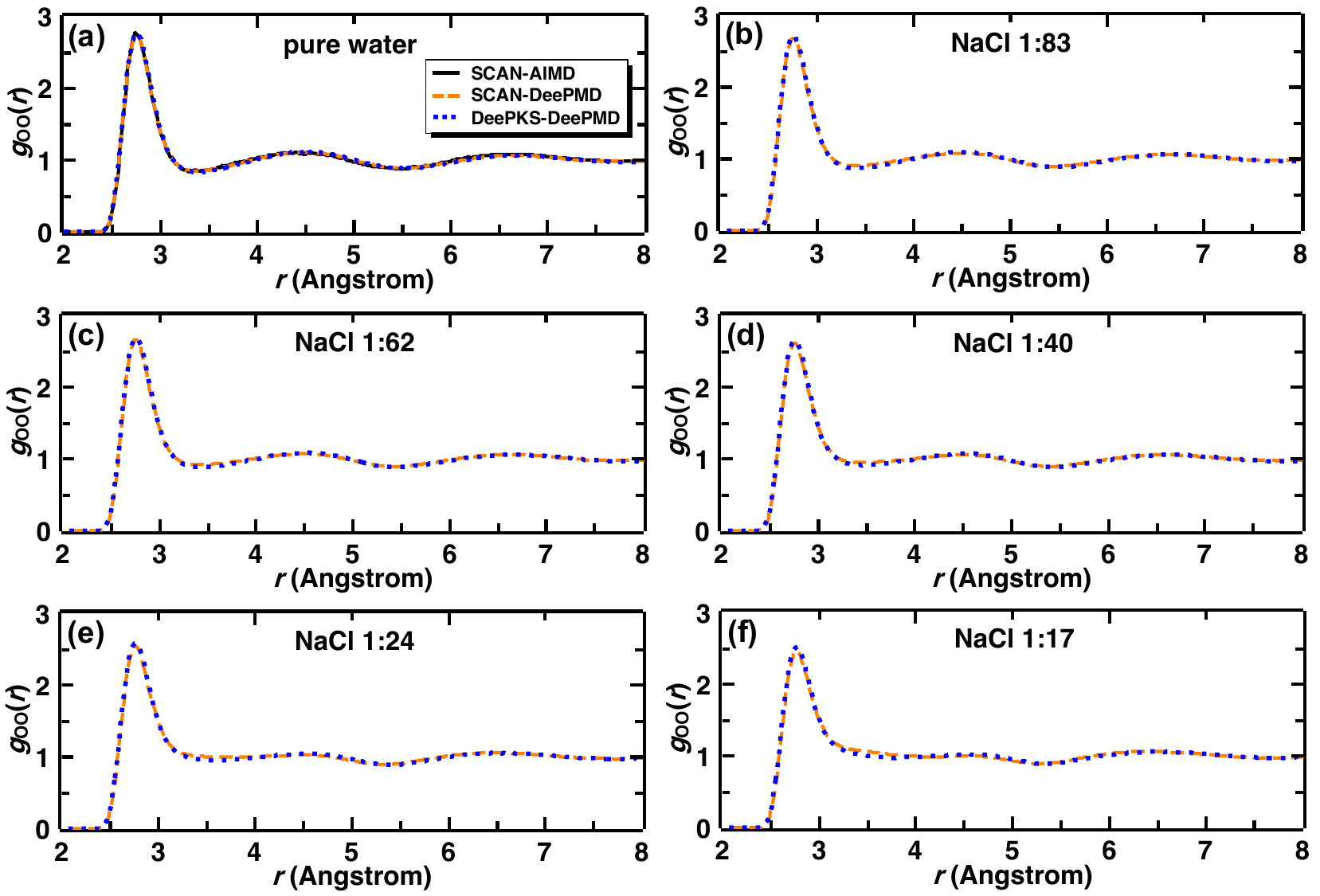}
    \caption{Oxygen-oxygen RDF [$g_{\scriptsize{\textrm{OO}}}(r)$] for {\bf(a)} pure water and NaCl solutions with concentration {\bf(b)} 1:83, {\bf(c)} 1:62, {\bf{(d)}} 1:40, {\bf{(e)}} 1:24, and {\bf{(f)}} 1:17 given by DeePKS-DeePMD (blue dotted line) from this work and SCAN-DeePMD (orange dashed line) from Ref. \citenum{chunyi_nc}. SCAN-AIMD result for pure water reported in Ref. \citenum{chunyi_nc} is also displayed for comparison (black solid line in {\bf(a)}).}
    \label{fig::goo_npt}
\end{figure*}

\begin{figure*}[htp]
    \includegraphics[width=0.5\linewidth]{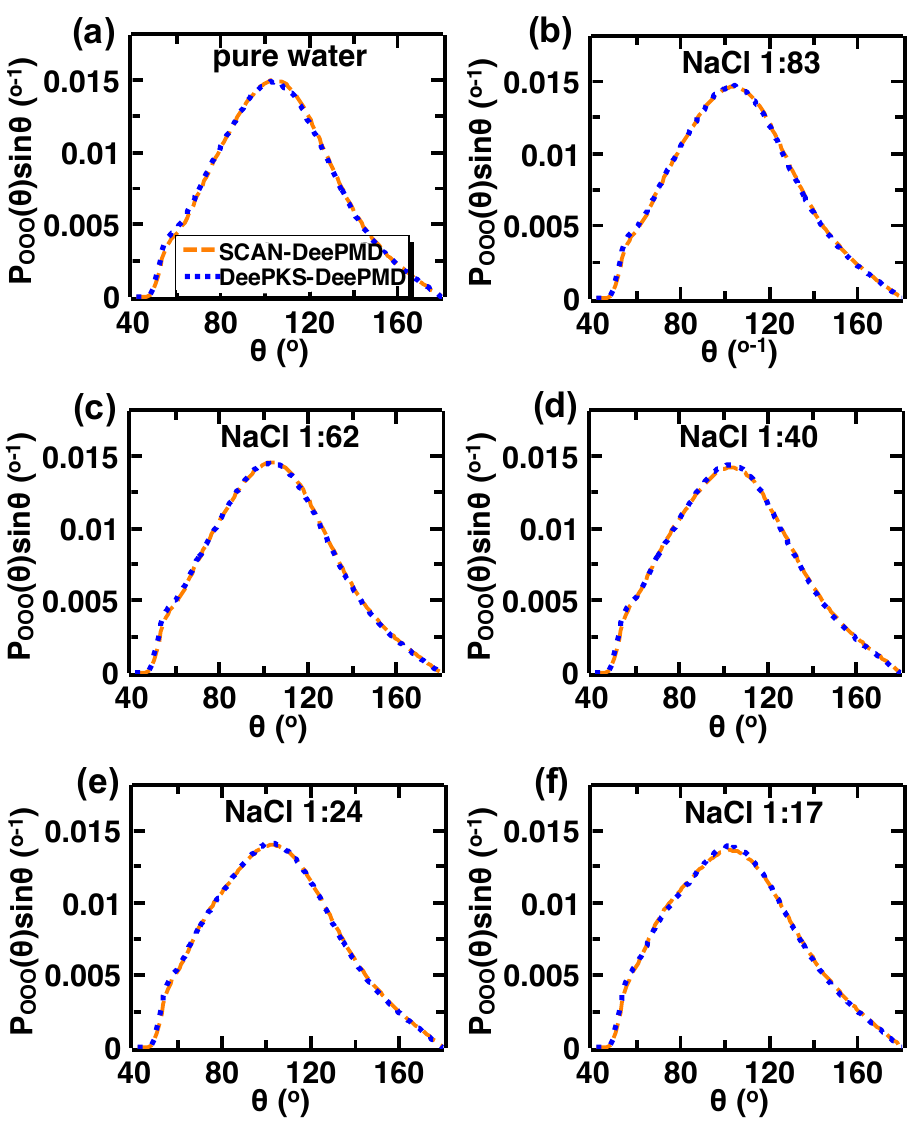}
    \caption{Bond angle distribution [$P_{\scriptsize{\textrm{OOO}}}(\theta)$] for {\bf(a)} pure water and NaCl solutions with concentration {\bf(b)} 1:83, {\bf(c)} 1:62, {\bf{(d)}} 1:40, {\bf{(e)}} 1:24, and {\bf{(f)}} 1:17 given by DeePKS-DeePMD (blue dotted line) from this work and SCAN-DeePMD (orange dashed line) from Ref. \citenum{chunyi_nc} }
    \label{fig::poo_npt}
\end{figure*}

It can be seen in Fig. \ref{fig::goo_npt}, Fig S2, and Fig S3 that for pure water and each investigated concentration, the oxygen-oxygen, oxygen-sodium,and oxygen-chloride RDFs predicted by DeePKS-DeePMD are nearly coincident with the SCAN-DeePMD results. 
As clearly shown in Fig S1, as the concentration increases, the population of interstitial water between the first and second peaks increases and the third coordination shell moves inwards with a diminishing feature of the second coordination shell, which is evinced by the pressure-like effect of salt water in the reciprocal space. 
The outward moving trend of the second shell with the increase of the concentration, which is highlighted in Ref. \citenum{chunyi_nc} as the major difference between the structures of salt and high-pressure water, is also successfully captured via DeePKS-DeePMD simulation as shown in Fig. S1. 
The ion-ion RDFs, which require considerably long simulation time to converge due to rather few ion pairs, are also calculated with the DeePKS-DeePMD trajectories. 
According to Fig. S4, $g_{\scriptsize{\textrm{NaNa}}}(r)$ and $g_{\scriptsize{\textrm{ClCl}}}(r)$ predicted via DeeKS-DeePMD are qualitatively in line with the SCAN-DeePMD results, with slight deviations that are presumably caused by the insufficiency of the simulation time. 
The bond angle distributions for pure and salt water with various concentrations are also explored (Fig. \ref{fig::poo_npt}) and delicate consistency is observed between DeePKS-DeePMD and SCAN-DeePMD results. 
With the increase of NaCl concentration, $P_{\scriptsize{\textrm{OOO}}}(\theta)$ shifts towards smaller angles from a tetrahedral distribution, which is mainly induced by the distribution of the first solvation shells of Na$^+$ as elucidated in Ref. \citenum{chunyi_nc}. 
The calculated bulk densities via DeePKS-DeePMD for NaCl solutions also closely match with those predicted via SCAN-DeePMD as shown in Fig. \ref{fig::density}(b).
The slight discrepancies at high concentration are presumably due to the fact that the SCAN stress is not included as a label during the DeePKS training process.
We shall leave this to our future investigations.

\section{Conclusion and Outlook}
\label{sec::conclusion}
In this work, we have bridged expensive \wh{high-level} $ab\ initio$ calculations and deep neural network potentials with the DeePKS model implemented in the open-source DFT software ABACUS. 
With less than two hundred frames of the training set labeled by \wh{hybrid meta-GGA or} meta-GGA functionals, we have shown that the GGA-based DeePKS model is able to quantitatively reproduce the target energies and forces for pure and salt water systems with \QO{orders} of magnitude savings in time \QO{(depending on the choice of the target method)}. 
The trained DeePKS model has then been applied in the DeePMD simulations for pure and salt water, i.e., two prototypical systems that are known to be poorly described via GGA functionals. The resulting structural and dynamical properties are in excellent agreement with the previously reported data obtained via \wh{hybrid meta-GGA or} meta-GGA AIMD and DeePMD methods, which underlines the reliability of the DeePKS model in connection with the DeePMD simulation. 

With the fully open-source implementation of the DeePKS+ABACUS methodology, we are expecting the spring up of extensive applications that require both high accuracy and computational efficiency. 
It is worth mentioning that even though the DeePKS model is trained on one or two specific periodic systems in this work, its transferability and generalizability should not be disregarded. 
Looking forward, it would be intriguing to develop the DeePKS model that is applicable to a class of systems such as electrolytes and inorganic semiconductors, enabling a generally accurate description which is hitherto challenging on such systems due to limited computational resources. 

\section{Acknowledgments}
The work of M.C. was supported by the National Science Foundation of China under Grant No. 12122401, 12074007, and 12135002.
The work of H.W.~was supported by the National Science Foundation of China under Grant No.11871110 and 12122103.
The work of C.Z. and X.W. were supported by National Science Foundation through Award No. DMR-2053195.
The work of Y.C. was supported by the Computational Chemical Sciences Center: Chemistry in Solution and at Interfaces (CSI) funded by DOE Award DE-SC0019394 and a gift from iFlytek to Princeton University.



\providecommand{\latin}[1]{#1}
\makeatletter
\providecommand{\doi}
  {\begingroup\let\do\@makeother\dospecials
  \catcode`\{=1 \catcode`\}=2 \doi@aux}
\providecommand{\doi@aux}[1]{\endgroup\texttt{#1}}
\makeatother
\providecommand*\mcitethebibliography{\thebibliography}
\csname @ifundefined\endcsname{endmcitethebibliography}
  {\let\endmcitethebibliography\endthebibliography}{}

\end{document}